\documentclass[twocolumn,showpacs,preprintnumbers,amsmath,amssymb]{revtex4}

\usepackage{graphicx}
\usepackage{dcolumn}
\usepackage{bm}
\usepackage{color}
\newcount\hours
\newcount\minutes   
\newcommand{\timeofday}{ 
\hours=\time
\minutes=\hours
\divide\hours by60
\multiply\hours by60
\advance\minutes by-\hours
\divide\hours by60
\ifnum\hours>9\else0\fi\the\hours:\ifnum\minutes>9\else
0\fi\the\minutes}

\begin{document}

\preprint{APS/123-QED}

\title{Ion acceleration processes at reforming collisionless shocks}

\author{R. E. Lee}
\email[]{leer@astro.warwick.ac.uk}
\author{S. C. Chapman}
\altaffiliation[Also at ]{Radcliffe Institute for Advanced
Study, Harvard University, USA}
\affiliation{Space and Astrophysics Group, Physics Department, University of
Warwick, Coventry, CV4 7AL, UK}
\author{R. O. Dendy}
\altaffiliation[Also at ]{Space and Astrophysics Group,
  Physics Department, University of Warwick, Coventry, CV4 7AL, UK }
\affiliation{UKAEA Culham Division, Culham Science Centre, Abingdon,
  Oxfordshire, OX14 3DB, UK} 

\date{\today \timeofday}

\begin{abstract}
The identification of pre-acceleration mechanisms for cosmic ray ions in
supernova remnant shocks is an important problem in astrophysics.
Recent particle-in-cell (PIC) shock simulations have shown that
inclusion of the full electron kinetics yields non-time-stationary
solutions, in contrast to previous hybrid (kinetic ions, fluid
electrons) simulations.  Here, by running a PIC code at high phase space
resolution, ion acceleration mechanisms associated with the time
dependence of a supercritical collisionless perpendicular shock are
examined. In particular the components of
$\int \mathbf{F} \cdot \mathbf{v} dt$ are analysed along trajectories
for ions that reach both high and low energies. Selection mechanisms for
the ions that reach high energies are also examined. In contrast to
quasi-stationary shock solutions, the suprathermal protons are selected
from the background population on the basis of the time at which they
arrive at the shock, and thus are generated in bursts.  
\end{abstract}


\maketitle

\section{INTRODUCTION}


Understanding the initial acceleration mechanisms for Galactic cosmic
rays remains an  outstanding problem in astrophysics. From energy balance
considerations, supernova remnants
(SNRs) provide the most
likely kinetic energy source to sustain the cosmic ray population. 
The local acceleration of electrons has been indirectly observed at the
expanding shock front of SNRs (see, for example, Ref.~\cite{koyama1995}).
However protons form the majority constituent of Galactic cosmic rays,
and  until recently observational evidence to link SNRs to local ion
acceleration has been lacking. X-ray and $\gamma$-ray data from
supernova remnant RX
J1713.7-3946 \cite{enomoto2002} show energy
spectra that can only be explained by accelerated ions.  Several
mechanisms are postulated to accelerate particles at SNR shocks.  Fermi
acceleration \citep{fermi1949}, which arises as a particle repeatedly
scatters off turbulent structures on either side of the shock, is in
principle capable of accelerating ions to high energies
\cite{bell1978}. However, to work effectively an initial suprathermal
population is required so that particles can re-cross the shock
front \cite{jokipii1987}. The identification and analysis of pre-acceleration
mechanisms that can select and initiate the energisation of completely
non-relativistic ions at SNR shocks from the background plasma is the
subject of this paper. 

The Rankine-Hugoniot relations \citep{tidmanandkrall} can be used to
determine the discontinuity in bulk plasma parameters across a
collisionless shock; that is, a shock where the particle mean free path
is much greater than length scales of interest. These relations are
derived by applying the magnetohydrodynamic (MHD) conservation equations
in the far upstream and downstream limits, away from the shock. Further
conditions are imposed by the fact that a shock must also increase
entropy, so that no subsonic flow can spontaneously become supersonic.
For an Alfv\'{e}nic Mach number $M_A \gtrsim 3$, the shock is
supercritical in that the increase in entropy, and in ion heating,
required by the Rankine-Hugoniot relations is achieved via the ion
kinetics; at least in part, by reflection of a fraction of upstream ions
at the shock. The generic supercritical,
quasi-perpendicular, 
collisionless shock in which ions reflect and gyrate in a foot region
upstream has been suggested by hybrid (particle ions, fluid electrons)
simulations (see, for example,
Refs.~\cite{quest1985,quest1986,burgess1989}), and confirmed by in-situ
observations of the Earth's bow shock \cite{sckopke1983}.

To study the acceleration of ions and electrons, a fully kinetic treatment
is in principle necessary for both species, and this can be closely
approximated by particle-in-cell (PIC) techniques. Physical mechanisms
operating on electron kinetic lengthscales and timescales are significant
both for aspects of macroscopic structure (for example, the shock ramp
width scales as $c/\omega_{pe}$), and for microscopic processes affecting the
ions (such as caviton formation and dissolution). Whether such effects are
important in any given scenario can be estimated, to some extent, by
analytical means, as we discuss in detail below. Importantly, however, it
is known that inclusion of the full electron kinetics can significantly
alter the dynamics of the shock. For example, hybrid simulations for
certain parameters \cite{quest1986,burgess1989} produce time-stationary
shock solutions, whereas for the same parameters PIC simulations reveal a
dynamic, reforming, shock structure. Furthermore the extent
to which an individual ion responds to phenomena on electron kinetic scales
must depend on that ion's cyclotron radius, and hence its energy. It
follows that for studies of ion acceleration at shocks, as in the
present paper, retention of full electron kinetics is desirable in order to
resolve fully the shock dynamics (see also
Refs.~\cite{scholer2003,lee2004}) and the ion
dynamics. We have previously presented  results of PIC code simulations
\cite{schmitz2002a,schmitz2002b,lee2004} that have high resolution in
real space and phase space, over relatively long run times, for
parameters relevant to shocks at supernova remnants. Our most recent
results \cite{lee2004} show that the time-dependent electromagnetic
fields at the reforming shock can accelerate inflow ions from background
to suprathermal energies. This provides a source population which may
subsequently be accelerated to produce high energy cosmic rays. 


In the present paper,
we focus on the specific nature of this ion acceleration
mechanism. This requires careful examination of the physics of the
interaction between particles and fields as they evolve over time. We
first introduce a methodology for simplifying the raw data, obtained on
spatio-temporal scales spanning those of the electrons and ions, into
data  suitable for examining events
on the spatio-temporal scales of interest for ion acceleration. We then
examine the detailed dynamics of the ion interactions with the shock
front, including a comparison between ions that eventually reach the
highest and lowest energies downstream. The time at which particles are
incident on the temporally evolving shock structure is found to be a key
discriminant in the subsequent energisation.

\section{SIMULATION METHOD}

The technical basis of the simulations was recently reported in
Ref.~\cite{lee2004}; let us reiterate briefly for completeness.
We use a relativistic electromagnetic particle-in-cell (PIC) code to
simulate the structure and evolution of a supercritical, collisionless,
perpendicular magnetosonic shock.  In a PIC simulation the distribution
functions of all particle species are represented by computational
super-particles, whilst the electromagnetic fields are defined on a
spatial grid. Particle trajectories are evolved from the fields via the
Lorentz force, then the fields are evolved from the new particle
positions via Maxwell's equations \cite{birdsallandlangdon}.  The code
we use to simulate the shock is based on that described in
Ref.\cite{devine1995}, and has been used recently to examine electron and
ion acceleration in SNR shocks
\cite{dieckmann2000,schmitz2002a,schmitz2002b,lee2004}. All vector
fields, bulk
plasma properties and particle velocities are functions of one spatial
co-ordinate ($x$), and time. This simplification enables detailed phase
space resolution for relatively long run times, however it constrains
magnetic fields: since $\nabla \cdot \mathbf{B} = 0$ we have $B_x =$
constant, taken as zero here in strict
perpendicular geometry.  PIC
simulations in two spatial dimensions (see, for
example, Ref. \cite{lembege1992}) that relax this constraint yield
overall shock dynamics that are consistent with the results seen here. 

We present results from simulations of a perpendicular shock
propagating into a magnetic field ($B_{z,1}$) of $1 \times 10^{-7}$
Tesla, a value consistent with those expected at supernova remnants
\cite{ellison1991}. The ratio of electron plasma frequency
to electron cyclotron frequency $\omega_{pe} / \omega_{ce} = 20$, and
the upstream ratio of plasma thermal pressure to magnetic field
pressure, $\beta = 0.15$. The shock has an Alfv\'{e}nic Mach number
($M_A$) of 10.5, and the simulation mass ratio for ions and electrons
$M_R = m_i / m_e = 20$, in common with Refs.\cite{shimada2000,schmitz2002a,
schmitz2002b,lee2004}.  This reduced mass ratio
is necessary to enable ion and electron time scales to be captured
within the same simulation, with reasonable computational overheads.
Previous PIC simulations for physical, and a range of non-physical, $m_i
/ m_e$ show a
variety of kinetic instabilities in the foot region
\cite{lembege2002,scholer2003}. Here we find that the ion dynamics
are insensitive to structures on electron scales, associated with these
instabilities. 

We also follow Refs.~\cite{shimada2000,schmitz2002a,schmitz2002b,lee2004}
in using the piston method (see, for example, Ref.~\cite{burgess1989},
and references therein) to set up the shock.  Particles are injected on
the left hand side of the simulation box with a drift speed $v_{inj}$,
modified by a small random velocity drawn from a thermal distribution,
characterised by
$u_{therm}$. At the particle injection boundary, the magnetic field
($B_{z,1}$) is constant and the electric field ($E_{y,1}$) is calculated
self consistently. The right hand boundary is taken to be a perfectly
conducting, perfectly reflecting wall. Particles reflect off this
boundary, and a shock then forms, propagating to the left through the
simulation box; sufficient time is allowed for the shock to propagate
sufficiently far upstream that the boundary conditions are no longer
important.  The size of a grid cell is defined as a Debye length
($\lambda_D$), and the time step is set to less than $\lambda_D / c$, for
numerical stability reasons \cite{hockneyandeastwood}. To enable the
shock and particle dynamics to be
followed over extended time-scales, whilst retaining high particle
density, a simple shock following algorithm is implemented.
This holds the peak in magnetic field at $8 \lambda_{ci}$ from the
left-hand boundary (for details see Ref.~\cite{schmitz2002a}, Appendix
A). This
distance is chosen so that no particles that are reflected off
the shock subsequently reach the upstream boundary, whilst it
leaves a significant region of the simulation box (around $
20 \lambda_{ci}$) downstream.

\section{RESULTS}

Full simulation of the  non-time-stationary features of a
collisionless shock requires the retention of electron dynamics; see,
for example Ref.~\cite{quest1986}.
However, resolving features on electron scales also introduces processes
that do not couple strongly to the processes that operate on ion scales,
which are the focus of the present paper. For example, the observed electron 
scale electrostatic solitary wave structures can lead to
electron acceleration \cite{schmitz2002a}, but do not significantly
affect the ions. As an aid to interpreting the interactions occurring
within a PIC 
simulation that give rise to ion acceleration, we present a
framework which distinguishes structure and dynamics on electron kinetic
scales from those relevant to ions. The ion trajectories that we present
here are
however all evolved self-consistently within the full
electro-magnetic fields of the PIC simulation

\subsection{Electric Potential on Ion Spatio-Temporal Scales \label{sec:pot}}

Resolving the full electron and ion kinetics in the PIC simulation
establishes two distinct spatio-temporal scales on which physical
processes can occur.  On sufficiently fast time scales and short length
scales there are, for example, plasma oscillations that lead to
fluctuations in charge density. However, on longer spatio-temporal
scales the plasma is quasi-neutral but still supports bulk electric
fields.  We can obtain an expression for these bulk fields by treating
the plasma as two fluids, ions and electrons (for a more general
multi-fluid treatment see, for example
Refs.~\cite{chapman1986,chapman1987,leroy1983}),  governed by the momentum
equations
\begin{eqnarray}
n_e m_e \frac{D\mathbf{v}_e}{Dt} &=& -e n_e (\mathbf{E} + \mathbf{v}_e
\mathbf{\wedge} 
\mathbf{B}) - \nabla P_e - n_e \nu_{ei} \mathbf{v}_e, \label{eqn:efluid}\\
n_i m_i \frac{D\mathbf{v}_i}{Dt} &=& q_i n_i (\mathbf{E} + \mathbf{v}_i
\mathbf{\wedge} 
\mathbf{B}) - \nabla P_i  - n_i \nu_{ie} \mathbf{v}_i \label{eqn:ifluid}.
\end{eqnarray}
The final terms in Eqs.~(\ref{eqn:efluid}) and (\ref{eqn:ifluid})
represent momentum transfer between species via forces not included in
the macroscopic fields. Here, we can assume that on the time-scale of ion
interaction with the shock, these terms are negligible for the ions.

We wish to consider space and time varying electro-magnetic fields that
only affect 
the ions, so that on ion scales we can take the limit in which the electrons
respond instantaneously as a charge neutralising fluid. This implies a 
vanishing electron inertial term,  the ``massless
electron fluid''  limit often used in hybrid codes:
\begin{equation}
m_e \frac{D\mathbf{v}_e}{Dt} \rightarrow 0.
\end{equation}
We neglect the electron pressure gradient because it is significant on
electron, rather than ion scales, however we anticipate that this
approximation will be least reliable in the shock ramp.

We can relate $\mathbf{v}_e$ directly to $\mathbf{v}_i$  via the current.
On the spatio-temporal scales on which the electron-proton ($q_i = e$)
plasma is quasi-neutral ($n_i \simeq n_e = n$),
\begin{equation}
 \mathbf{J} \simeq en(\mathbf{v}_i - \mathbf{v}_e).\label{eqn:vesubs}
\end{equation}
Substitution for $\mathbf{v}_e$ from Eq.~(\ref{eqn:vesubs}) into
Eq.~(\ref{eqn:efluid}) then gives
\begin{equation}
0 \simeq  \mathbf{E} + \left[\mathbf{v}_i -
\frac{\mathbf{J}}{en} \right] \mathbf{\wedge} \mathbf{B}.  
\label{eqn:efluidsubs}
\end{equation}
Consistent with this low frequency approximation we neglect the
displacement current (the non-radiative limit), giving Amp\`{e}re's law
\begin{equation}
 \nabla \wedge \mathbf{B} = \mu_0 \mathbf{J}.
\end{equation}
This implies the standard expression
\begin{equation}
\mathbf{J} \wedge \mathbf{B} =  - \frac{1}{\mu_0} \left[ \frac{\nabla
  B^2}{2} - \left( \mathbf{B} \cdot \mathbf{\nabla}
  \right)\mathbf{B} \right]. \label{eqn:jcrossb}
\end{equation}
Together, Eqs.~(\ref{eqn:efluidsubs}) and (\ref{eqn:jcrossb}) give
\begin{equation}
0  \simeq  \mathbf{E} + \mathbf{v}_i \wedge \mathbf{B} +  
  \frac{1}{\mu_0en} \left[ \frac{\nabla
  B^2}{2} - \left( \mathbf{B} \cdot \mathbf{\nabla}
  \right)\mathbf{B} \right],  \label{eqn:estuff}
\end{equation}
which can then be rearranged to yield $\mathbf{E}$. 

In the one dimensional geometry of our simulation Eq.~(\ref{eqn:estuff})
can be simplified by noting $\nabla \equiv (\partial_x , 0 , 0)$, thus
$(\mathbf{B} \cdot \mathbf{\nabla} ) \mathbf{B} = 0$.  Further
simplification arises if we note that generally in our simulations,
$v_z << v_y$, thus $(v_{i,y}B_z - v_{i,z}B_y ) \simeq
v_{i,y}B_z$. Rearranging a simplified Eq.~(\ref{eqn:estuff}) then gives
\begin{eqnarray}
E_{x,i}  &\simeq&  - \frac{1}{e n \mu_0} \frac{\partial
  (B_z^2 / 2 + B_y^2 / 2)}{\partial x} -  v_{i,y}B_z, \label{eqn:ex}\\
E_{y,i} &\simeq& v_{i,x} B_z. \label{eqn:ey}
\end{eqnarray}
Here $E_{x,i}$ is the $x$ component, and $E_{y,i}$ the $y$ component, of the
electric field on the slow,  ion spatio-temporal scales on which the
plasma is quasi-neutral.  

Substitution of our simplified, ion scale, electric field
from Eq.~(\ref{eqn:estuff}), into the ion force equation
Eq.~(\ref{eqn:ifluid}), leads to an expression whose $x$ component is
\begin{equation}
\left. nm_i \frac{D\mathbf{v}_i}{Dt}\right|_x \simeq - \frac{1}{\mu_0}
\frac{\partial (B_z^2/2 + B_y^2/2)}{\partial x} - \frac{\partial
  P_i}{\partial x}. \label{eqn:ifluid2}
\end{equation}

It follows from Eq.~(\ref{eqn:ifluid2}) that the bulk force on the ion
fluid is due to
gradients in magnetic and plasma pressure. The potentials (which act on
individual particles) follow from Eqs.~(\ref{eqn:ex}) and (\ref{eqn:ey}).

\begin{figure}
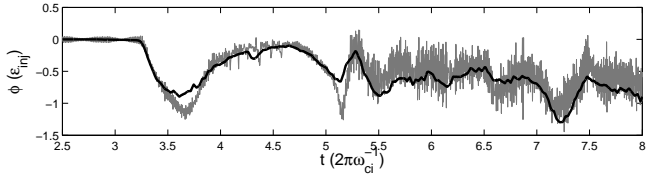

\caption{$\phi = \int E_x dx$ from PIC code (gray), and $\phi_i = \int E_{x,i} dx$
  calculated from Eq.~(\ref{eqn:ex}) (black) along the trajectory of a high
  energy ion. In calculating $E_{x,i}$ we compute the ion flow
  velocity from the mean velocity of all ions within $0.02 \lambda_{ci}
  \simeq \lambda_{ce}/2 \simeq 10$ grid cells of the particle position.} 
\label{fig:pot}
\end{figure}

Figure~\ref{fig:pot} demonstrates the extent to which this approximate
analytical treatment provides a guide to the ion behaviour that is
calculated from first principles in the PIC code. It compares the time
evolution of the potential, $\phi = \int E_x dx$, obtained directly from the
PIC code, to that calculated on ion scales, $\phi_i = \int E_{x,i} dx$, using
Eq.~(\ref{eqn:ex}). The path chosen for the spatial integration is that
of an ion that reaches a high energy on leaving the shock
front. Fig.~\ref{fig:pot} demonstrates that $E_{x,i}$ defined in
Eq.~(\ref{eqn:ex}) is a useful guide, and hence the analysis above
captures much of the key physics. The ion scale bulk potential
essentially averages over the small scale fluctuations of the ``raw''
potential. We can see from Fig.~\ref{fig:pot}, however, that the average
values of $\phi_i$ depart from that of the full potential $\phi$ where
the ion interacts with the shock ramp: first during a reflection at $t = 3.5 -
3.7$, and during a subsequent transmission to downstream at $t = 5 - 5.2
\times 2\pi\omega_{ci}^{-1}$. In the discussion below, we will calculate
the ion energetics from the full electromagnetic fields of the PIC simulation.


\subsection{Ion Acceleration \label{sec:id}}

To study the physical processes that cause ion acceleration, we
evaluate the changes in kinetic energy of ions during their
interaction with the shock. Here the ion Lorentz factor $\gamma
\sim 1$, therefore we can neglect relativistic effects.  We have
\begin{equation}
\mathbf{F} = m \frac{d \mathbf{v}}{dt} = q(\mathbf{E} + \mathbf{v \wedge B})
\end{equation}
where, in our collisionless plasma, $\mathbf{E}$ and $\mathbf{B}$ in the
Lorentz force refer to the fields in the PIC simulation. Thus
\begin{equation}
\mathbf{F \cdot v} = \frac{d}{dt} \left(\frac{1}{2}mv^2 \right)
= q\mathbf{E \cdot v}  
\end{equation}
Integration along a computed ion trajectory then implies that the
kinetic energy acquired is: 
\begin{equation}
\frac{1}{2}mv^2 = q \int_{trajectory}
\mathbf{E} \cdot \mathbf{v} dt. \label{eqn:ke} 
\end{equation}

\subsubsection{Highly energetic ions\label{sec:highE}}
Previous PIC simulations \cite{lee2004} have shown that the
downstream proton population has a continuous distribution of
energies from zero up to $\sim 6$ times the ion injection energy, ${\cal
E}_{inj} = \frac{1}{2}m v_{inj}^2$, in the frame in which the
downstream plasma is at rest. We now examine the dynamics of these ions
in more detail. Figure~\ref{fig:HighE-fdotv} presents the results of
evaluating Eq.~(\ref{eqn:ke}) for a 
selected group of protons that become highly energised. The top panel
(panel 1) displays the 
kinetic energy over time, calculated from $\int q\mathbf{E} \cdot
\mathbf{v} dt$ along the particle trajectory. Panel 2 shows only the
$x$-component, $\int qE_x v_x dt$ normal to the shock, and panel 3 shows
only the $y$-component along the shock front. The $z$-component is
omitted as it remains identically zero, due to the configuration of the
simulation domain. Panel 4 displays the potential, $\phi = \int E_x dx$,
computed directly from the PIC code, at the $x$-position of the ions at
the current time. Panel 5 shows the $y$-positions. In the lowest panel
(panel 6) the $x$-positions are shown in relation to the
spatio-temporally evolving potential
structure on ion scales, $\phi_i = \int E_{x,i} dx$ computed using
Eq.~(\ref{eqn:ex}). Comparison of this panel with Fig.~8 of Ref.~\cite{lee2004}
shows that $\phi_i$ captures the qualitative features of the electromagnetic
fields. To complement this information, Fig.~\ref{fig:1p-xy}
shows the trajectory of a high energy ion in the $x$-$y$ plane (in the
simulations we evolve the three components of the particle velocity
$\mathbf{v}(x,t)$, and these can be integrated to provide a trajectory in
three dimensional configuration space). As with all results in this
paper, data is presented in the downstream rest frame, and has been
obtained from a segment of the simulation when the shock is propagating
independently of the boundary conditions. Units are normalised to the
upstream ion parameters, that is, $\lambda_{ci}$ the upstream ion
cyclotron radius, and $\omega_{ci}$ the upstream ion cyclotron
frequency.



Figure~\ref{fig:HighE-fdotv} shows that the ions that become highly
energised remain close in phase space throughout their interactions with
the shock. After passing through the shock, local fluctuations in the
fields lead to some
divergence in the $y$-component of $\int q\mathbf{E} \cdot \mathbf{v} dt$
and the $y$-position. Panel 6 shows the shock propagating in the
negative $x$-direction over time, while undergoing reformation
cycles. The size and depth of the potential well varies over the course
of a reformation cycle, on a time scale comparable to the local ion
cyclotron period, as discussed in detail by \citet{lee2004}.

If we follow the path through the shock region of an individual ion that
eventually reaches high energy, a series of events occurs.  The ion is
initially co-moving with the plasma at the inflow speed. This
corresponds to the linear increase in $x$-position (panel 6), with no
translation in the $y$-direction (panel 5 and Fig.~\ref{fig:1p-xy}), all
at the inflow energy (panel 1). It should be noted that in panel 1, the
energies are initially zero, because the start of the integration path
for $\int q\mathbf{E} \cdot \mathbf{v} dt$ is chosen to be just inside the
simulation domain, where the ions are already co-moving with the
plasma. This location is upstream of the shock foot, and the potential
in this region is therefore constant, with the only variation due to
high frequency fluctuations in $\mathbf{E}$.

After approximately $3.25$ ion cyclotron periods, the ion enters the
potential well upstream of the shock front, point {\it a} in
Fig.~\ref{fig:HighE-fdotv}. At this time, 
the shock is most fully formed, the shock jump is close to maximal, so
that the $\partial B^2 / \partial x$ term in the
potential from Eq.~(\ref{eqn:ex}) is close to maximal also.
The $x$-component of $\int q\mathbf{E} \cdot
\mathbf{v} dt$ shows a decrease in energy as the ion journeys further
into the well (panel 2), with a corresponding decrease in kinetic energy
(panel 1); this follows from a negative value for $E_x$. Between points
{\it a} and {\it b} there a is 
decrease in $x$-velocity as the ion gets closer to the shock and the
magnetic field increases, which is accompanied by a weak drift in $-y$,
see also Fig.~\ref{fig:1p-xy}. 

At the time corresponding to point {\it b}, the kinetic energy reaches a
minimum (panel 1), when the ion is near the deepest point in the
potential well (panels 4 and 6). The ion has now stopped its
progression toward the shock front (Fig.~\ref{fig:1p-xy}), and reflection
back into the upstream region has begun (panel 6). 

By time {\it c} the ion has begun to climb back out of the potential well,
away from the shock front (panel 4). After {\it c}, drift in $+y$ begins
as the ions move in $-x$ into the foot region, see also
Fig.~\ref{fig:1p-xy}. Meanwhile the 
$x$-component of $\int q\mathbf{E} \cdot \mathbf{v} dt$ starts to
increase rapidly (panel 2), due to the strength of $E_x$ (shown by the
gradient of $\phi$, panel 4).

By point {\it d} the particle has moved back to the extreme upstream
edge of the potential well. It then remains moving along a contour of
$\phi \simeq 0$ (panels 4 and 6) for a time approximately equal to one
upstream ion cyclotron period ($2 \pi \omega_{ci}^{-1}$). Between {\it
  d} and {\it  e} the value of $E_x$ 
local to the ion is lower, so that the associated energisation rate
is also less; however the positive
$y$-component of the gyromotion continues (panel 5 and
Fig.~\ref{fig:1p-xy}), and since the motional $E_y$ is positive, this
gives an energy gain 
in the $y$-component $\Delta {\cal E}_y = \Delta mv_y^2 / 2$ (panel
3). The gyromotion of the particle eventually leads to a positive
$x$-component of velocity (Fig.~\ref{fig:1p-xy}), so that at point {\it
e} the particle finally leaves the extreme upstream edge of the well and
passes through the potential well for a second time (panels 4 and 6).
This marks the end of a prolonged episode of energy gain, which now
stops as $y$-drift ceases (panels 1 and 3), and the particle settles
into its stable downstream gyromotion (Fig.~\ref{fig:1p-xy}). The ions
cross the saddle in the potential $\phi(x,t)$ (as shown in panel 6),
leading to a brief energy loss then gain via the $x$-component of $\int
q \mathbf{E} \cdot \mathbf{v} dt$ (panel 2), before the ion passes
thorough the shock front (point {\it f}, panels 3 and 6), and gyrates
away from the shock into the downstream region. The ion energy now
exceeds its initial value by a factor of approximately six.

In summary there are two stages of acceleration as shown in
Fig.~\ref{fig:1p-xy}: normal reflection from the temporarily
stationary shock front into the foot region, followed by energisation
during transverse drift across the shock front. 

\subsubsection{Weakly energised ions\label{sec:lowE}}

\begin{figure}[floatfix]
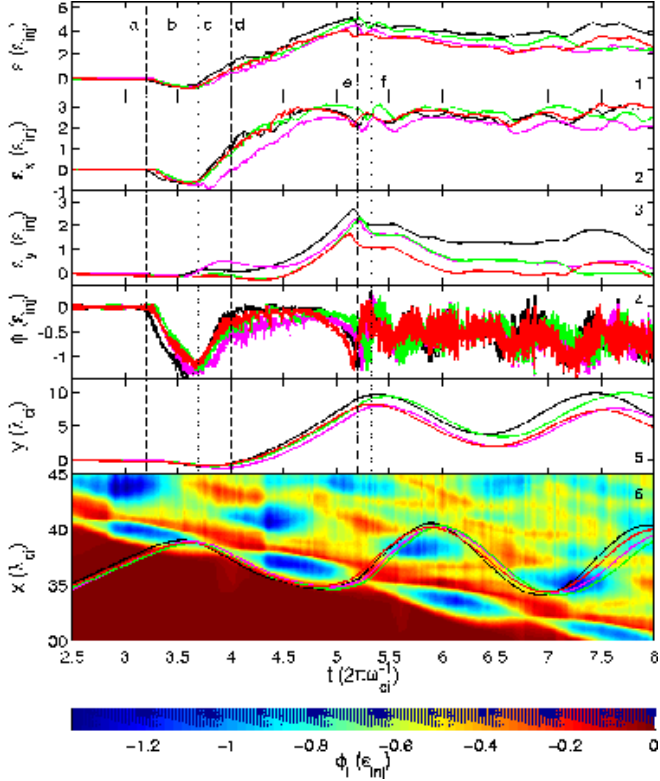

\caption{Trajectories for 4 ions that reach high energies. Panel
  1 shows $\int q\mathbf{E} \cdot \mathbf{v} dt$ along each
  trajectory. Panel 2 the $x$-component of $\int q\mathbf{E} \cdot
  \mathbf{v} dt$, panel 3 the $y$-component. The $z$-component is not
  shown as it remain identically 0. Panel 4 displays $\phi = \int E_x
  dx$. Panel 5, $y$-position and panel 6, $x$-position plotted over the
  potential on ion scales, derived from Eq.~(\ref{eqn:ex}), here the
  shock is propagating towards lower values of $x$ as $t$ increases. The
  vertical lines correspond to times of change during the black trajectory.
}
\label{fig:HighE-fdotv}
\end{figure}

\begin{figure}[floatfix]
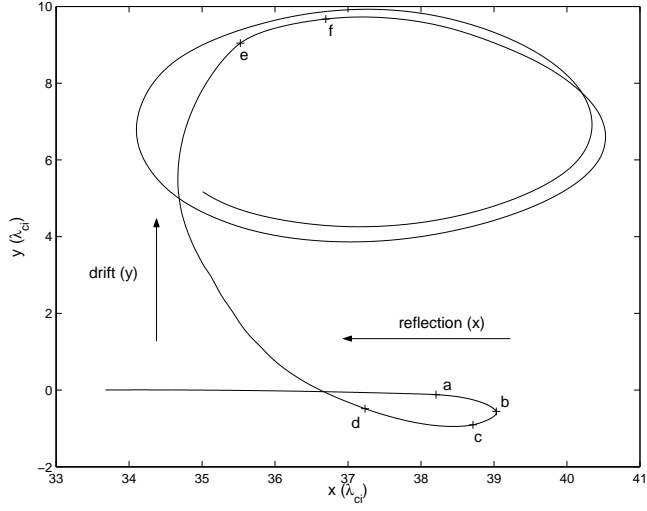

\caption{Position $x$ vs. $y$ for a high energy ion. This ion follows
  the black trajectory in Fig.~\ref{fig:HighE-fdotv} and the timing
  points {\it a-f} are indicated.}  
\label{fig:1p-xy}
\end{figure}

Further insight into the energisation process can be gained by comparing
trajectories for ions that become highly energised
(Fig.~\ref{fig:HighE-fdotv}), to those for a group
of ions that have only low energies ($\lesssim {\cal E}_{inj}$) on finally
entering the downstream region (Fig.~\ref{fig:LowE-fdotv}), and remain in
the core of the downstream particle distribution.



The trajectories for the ions that are not subsequently
energised are initially similar to those for the ions that eventually
reach the higher energies.  The primary difference is the timing of
their first encounter with the shock. We have identified two distinct
groups of low energy ions, and these are shown in
Fig.~\ref{fig:LowE-fdotv}. The first group arrives at the shock front
just as the 
shock is advancing ($t = 3.1 \times 2\pi\omega_{ci}^{-1}$),
and the second when the shock is decaying ($t = 4 \times
2\pi\omega_{ci}^{-1}$).

The reforming shock progresses upstream (downwards in panel 6 of
Figs.~\ref{fig:HighE-fdotv} and \ref{fig:LowE-fdotv}) in a stepwise
fashion. The first group of ions (upper trajectories in panel 6
of Fig.~\ref{fig:LowE-fdotv}) encounter the advancing shock when the
shock jump is sufficiently large to cause reflection (between points $A$
and $B$). Their trajectories up to this
point are akin to the trajectories up to point {\it b} in
Fig.~\ref{fig:HighE-fdotv} for the ions that have become highly energised: they
have entered the foot region (point $A$) and 
been deflected in the $-y$ direction, whilst losing energy via a
decrease in $\int qE_x v_x dt$. However, at this time the shock speed is
maximal, so their velocity component in $-x$ is smaller than that of the
shock itself, and the shock overtakes them. They then co-move with the
shock front for about an upstream ion cyclotron period, before moving
downstream.

On the other hand the second group of ions encounter the shock when the
potential is decaying (point $A'$). They then pass through the potential
well (point $B'$), and reach the shock without reflection, where their
$v_\perp$ increases (Fig.~\ref{fig:LowE-xy}) along with the bulk
$\mathbf{B}$ field downstream.
  
Regardless of their energisation or of the details of their dynamics,
the guiding center velocity of all ions goes to zero once they have
propagated sufficiently far downstream of the shock front. This is to be
expected, because the far downstream frame defines the rest frame of the
plasma, as noted previously (Figs.~\ref{fig:1p-xy} and
\ref{fig:LowE-xy}). 

Of further interest is the $y$-motion of the two groups of low energy
ions (panel 5 of Fig.~\ref{fig:LowE-fdotv}). Those that enter the shock
before the high energy ions have little movement of their gyrocenters in
$y$, but those that enter after the high energy ions, have a significant
$-y$ drift velocity; see also Fig.~\ref{fig:LowE-xy}. Both these patterns are
in contrast to the high 
energy ions that have gyrocenters that drift in the $+y$ direction
(panel 5 of Fig.~\ref{fig:HighE-fdotv}, and Fig.~\ref{fig:1p-xy}).

Finally we can compare the kinetic energy gain computed directly from
the electromagnetic fields of the PIC simulation, with that given by the
fields on ion scales estimated from Eqs.~(\ref{eqn:ex}) and (\ref{eqn:ey}). In
Fig.~\ref{fig:fdotvcompare} we plot the total kinetic energy as a
function of time along the trajectory of an ion that is reflected and
reaches suprathermal energy. This plot shows a close correspondence
between the two curves, suggesting that the fine structure on electron
scales does not affect the final energy gain of these ions. However, as
we have seen from Fig.~\ref{fig:pot}, there is a discrepancy between the
$x$-component of the electric field at the shock ramp obtained from the
simulation directly, and from Eq.~(\ref{eqn:ex}). It therefore follows
from Fig.~\ref{fig:fdotvcompare} that the value of the shock
ramp does not strongly affect the overall energy gain of
these ions. This is consistent with energisation being associated with
electromagnetic fields away from the ramp, the role of the
ramp potential being simply to reflect the ions. However, details of the
energetics of low energy ions that are not reflected, may depend on the
value of the shock ramp potential.


\subsection{Role of Shock Reformation}

Having examined the trajectories of ions that reach a variety of
energies, let us now examine the selection mechanisms that give rise to
different histories and energisation. For a time stationary shock,
\citeauthor{burgess1989}~\cite{burgess1989} examined the origins in phase
space of ions that eventually reach differing energies. Particles from
the extrema of the velocity space distribution upstream of the shock
were found to be 
preferentially reflected further upstream, and so energised to higher
energies; whereas ions from the core of the distribution passed through
the shock, moving
little or no distance upstream. To establish whether the same selection
mechanism is at work in our dynamic reforming shock, we have constructed in
Fig.~\ref{fig:burgess} a series of plots that may be compared with Fig.~1 of
Ref.~\cite{burgess1989}. Figure~\ref{fig:burgess} shows the ion phase
space ($v_x$ and $v_y$ vs. $x$) for groups of ions at differing initial
perpendicular velocities, at a time when the potential well is at its
narrowest, and the reformation cycle has just commenced, corresponding
to $t = 4 \times 2 \pi \omega_{ci}^{-1}$ in Fig.~\ref{fig:HighE-fdotv}.
In contrast to the results obtained in Ref.~\cite{burgess1989} for the
case of a time stationary shock, we find that at a reforming shock, ions
that are initially in the core of the distribution, as well as those
from the tails, are reflected back into the foot region. Also, the
distance travelled back into the foot region (and hence the energy
gained) appears independent of the initial perpendicular velocity of the
ions. Whether or not a given ion is reflected depends on its normal
velocity in comparison with the time-dependent shock potential. Thus
the timing at which ions arrive at the shock front determines their
final location in velocity space.


Overall, examination of the shock dynamics in relation to ion
trajectories shows that the ions that are ultimately highly energised
(Fig.~\ref{fig:HighE-fdotv}) reflect from the shock front just as it
becomes stationary, and pass through the foot region saddle in
$\phi(x,t)$ (panel 6). The ions that meet the shock
front prior to this (upper group in Fig.~\ref{fig:LowE-fdotv}) interact
with a shock front that is moving rapidly forward through the simulation
domain, and so do not gain sufficient velocity to outpace it. The ions
that interact later (lower group in Fig.~\ref{fig:LowE-fdotv}) meet a
weakening shock front with a wider potential well, so that they are not
reflected. In contrast to Fig.~\ref{fig:HighE-fdotv}, the ions in
Fig.~\ref{fig:LowE-fdotv} experience neither an $x$-energy gain on
moving back to the upstream side of the potential well, nor a sustained
period of $y$-energy gain as they subsequently co-move with the upstream
edge of the well.

The present results suggest that the time-evolving shock dynamics, and
in particular the timing of the interaction between ion and shock, govern
the selection process determining which ions undergo pre-acceleration
into a suprathermal population that may subsequently become cosmic rays.

\section{CONCLUSIONS}

We have examined in detail the dynamics of suprathermal ions generated
in PIC code simulations of quasi-perpendicular reforming
shocks. Importantly, this energisation is not found in stationary shock
solutions. We find that:
\begin{enumerate}
\item The shock structure reforms on a timescales of the order of the local ion
  cyclotron period. This is shown clearly if the electromagnetic fields
  are cast in the form of a potential, after removing small scale
  effects, to leave only terms relevant on ion spatio-temporal scales.
\item The time-dependence of the shock dominates the selection of which
  ions are accelerated to suprathermal energies. Ions that reach the
  shock when its ramp, and hence potential, are maximal, are reflected
  and subsequently gain energy by drifting in the time-dependent fields
  tangential to the shock front.
\item  This selection is in contrast to a time-stationary shock, where the
selection mechanism depends upon the initial ion velocity perpendicular
to the magnetic field, those  ions coming from the tails of the
distribution being preferentially reflected, and so energised.
\item These factors lead to high energy ion creation occurring in bursts.  
\end{enumerate}

The present simulations are conducted in a $1x3v$
geometry at a low $m_i/m_e$. Whilst the work in 
Ref. \cite{lembege1992} shows no major differences in  higher
spatial dimensions for PIC simulations, hybrid simulations in $2x3v$,
for longer run times,
show that the current that can now exist along the shock front can lead
to current-driven instabilities \cite{winske1988}. These instabilities
may act to change 
the shock structure in the $y$-direction, and so alter ion and electron
dynamics across the shock front, affecting both ion and electron acceleration.
Simulations with more realistic $m_i / m_e$ ratios
\cite{lembege2002,scholer2003} show alterations in the electron scale
physics in the foot region. However, we have shown that the electron
scale physics has little effect on ion spatio-temporal scales.  

The plasmas simulated here are pure hydrogen, in that
there are only two species, protons and electrons. Both species have a
Maxwellian distribution with the same temperature. The addition of
pickup ions to the simulation, for example in relation to the
heliospheric termination shock, where hybrid simulation have already
been carried out \cite{lipatov1999,lever2001}, would allow the
acceleration processes relevant to anomalous cosmic ray production to be
examined. This will be the subject of future work. 

The fundamental plasma physics processes underlying the ion acceleration
from background to suprathermal energies (10 to 20 MeV), reported in the
SNR shock simulations of Ref.~\cite{lee2004}, have been elucidated in
the present 
paper. Specifically, we have explained the role of, and interplay between,
the key elements anticipated at the end of the Appendix to
Ref.~\cite{lee2004}. We have shown that, while an
electron fluid approximation captures some of the key physics, the shock
reformation dynamics arising from our fully kinetic PIC treatment are
central to the ion acceleration mechanism. This work provides a clear first
principles explanation for the ion acceleration that
is observed in our simulations, which appears to be a strong candidate
injection mechanism for Galactic cosmic ray protons.

\begin{acknowledgments}
R.~E.~L. acknowledges a CASE Research Studentship from the UK Particle
Physics and Astronomy Research Council in association with UK Atomic
Energy Authority and a
Warwick Postgraduate Research Fellowship from the University of Warwick. This
work was also supported in part by the UK Engineering and Physical
Sciences Research Council. S.~C.~C. acknowledges the Radcliffe
Foundation, Harvard.
\end{acknowledgments}


\clearpage

\begin{figure}[floatfix]
\caption{Trajectories for four ions that remain at low energies when
  crossing the shock front. Panels and colors are as
  Fig.~\ref{fig:HighE-fdotv}. 
}
\label{fig:LowE-fdotv}
\end{figure}

\begin{figure}[floatfix]
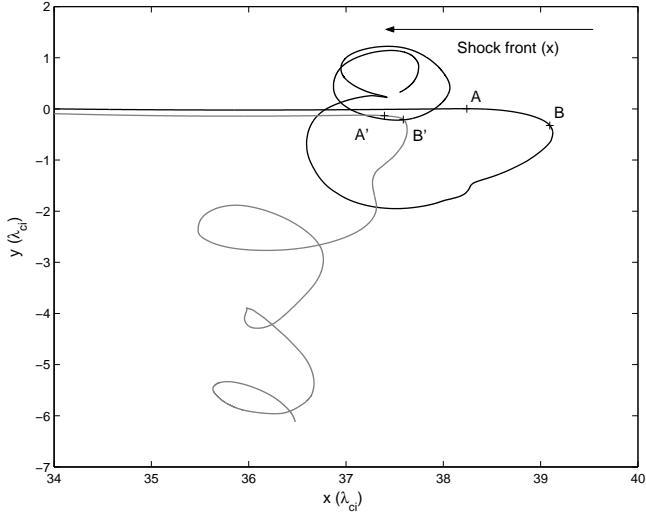

\caption{Position $x$ vs. $y$ for two low energy ions. The black and
  gray trajectories from Fig.~\ref{fig:LowE-fdotv} are represented here,
  with the timing points $A,B,A',B'$ indicated.}
\label{fig:LowE-xy}
\end{figure}

\begin{figure}[floatfix]
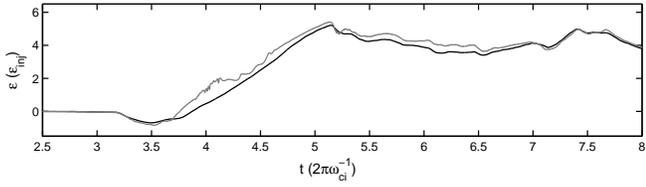

\caption{Kinetic energy calculated from Eq.~(\ref{eqn:ke}) using the
  PIC simulation $\mathbf{E}$ (gray) and that on ion scales $\mathbf{E}_i$
  (black) from Eqs.~(\ref{eqn:ex}) and (\ref{eqn:ey}), for
  the black ion in Figs.~\ref{fig:HighE-fdotv} and \ref{fig:1p-xy}. The bulk
  ion velocity $v_{ix,y}$ in Eqs.~(\ref{eqn:ex}) and (\ref{eqn:ey}) is
  calculated from the mean velocity of the ions within $10$ grid cells
  $\simeq 0.02\lambda_{ci} \simeq \lambda_{ce}/2$ of the particle position.}
\label{fig:fdotvcompare}
\end{figure}

\begin{figure}[floatfix]
\caption{Phase space plots of $v_x$ (left) and $v_y$ (right) vs. $x$ for
  groups of particles 
  from differing regions of initial velocity space, at the instant $t = 4
  \times 2 \pi \omega_{ci}^{-1}$.} 
\label{fig:burgess}
\end{figure}

\clearpage
\setcounter{figure}{0}
\begin{figure}
\noindent\includegraphics[width=\columnwidth]{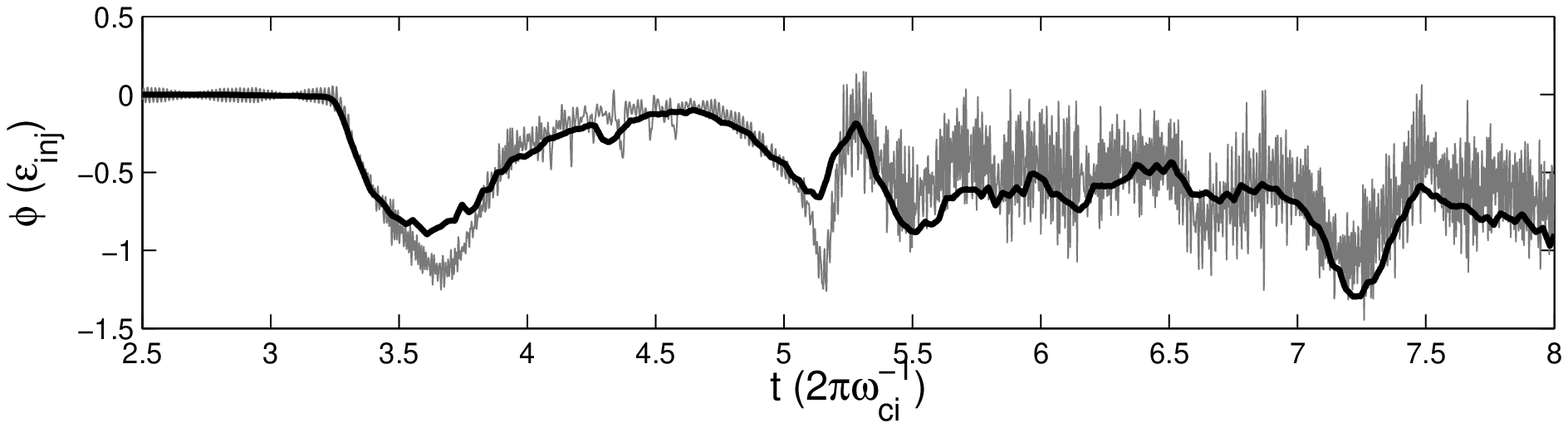}
\caption{$\phi = \int E_x dx$ from PIC code (gray), and $\phi_i = \int E_{x,i} dx$
  calculated from Eq.~(\ref{eqn:ex}) (black) along the trajectory of a high
  energy ion. In calculating $E_{x,i}$ we compute the ion flow
  velocity from the mean velocity of all ions within $0.02 \lambda_{ci}
  \simeq \lambda_{ce}/2 \simeq 10$ grid cells of the particle position.} 
\end{figure}

\begin{figure}[floatfix]
\noindent\includegraphics[width=\columnwidth]{037412php2.col}
\caption{Trajectories for 4 ions that reach high energies. Panel
  1 shows $\int q\mathbf{E} \cdot \mathbf{v} dt$ along each
  trajectory. Panel 2 the $x$-component of $\int q\mathbf{E} \cdot
  \mathbf{v} dt$, panel 3 the $y$-component. The $z$-component is not
  shown as it remain identically 0. Panel 4 displays $\phi = \int E_x
  dx$. Panel 5, $y$-position and panel 6, $x$-position plotted over the
  potential on ion scales, derived from Eq.~(\ref{eqn:ex}), here the
  shock is propagating towards lower values of $x$ as $t$ increases. The
  vertical lines correspond to times of change during the black trajectory.
}
\end{figure}
\begin{figure}[floatfix]
\noindent\includegraphics[width=\columnwidth]{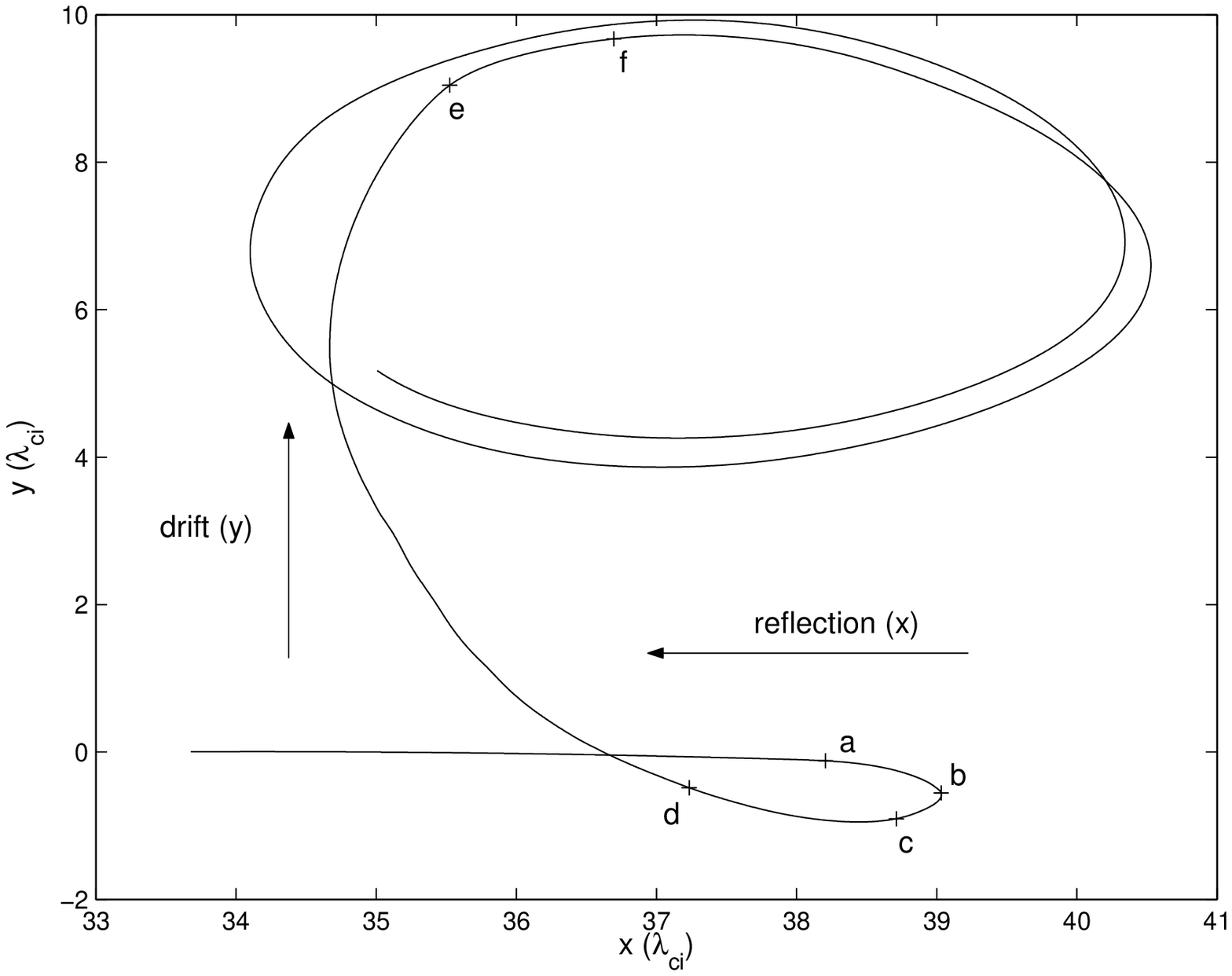}
\caption{Position $x$ vs. $y$ for a high energy ion. This ion follows
  the black trajectory in Fig.~\ref{fig:HighE-fdotv} and the timing
  points {\it a-f} are indicated.}  
\end{figure}

\begin{figure}[floatfix]
\noindent\includegraphics[width=\columnwidth]{037412php4.col}
\caption{Trajectories for four ions that remain at low energies when
  crossing the shock front. Panels and colors are as
  Fig.~\ref{fig:HighE-fdotv}. 
Black  vertical lines refer to the black trajectory, red lines to the red
  trajectory.
}
\end{figure}

\begin{figure}[floatfix]
\noindent\includegraphics[width=\columnwidth]{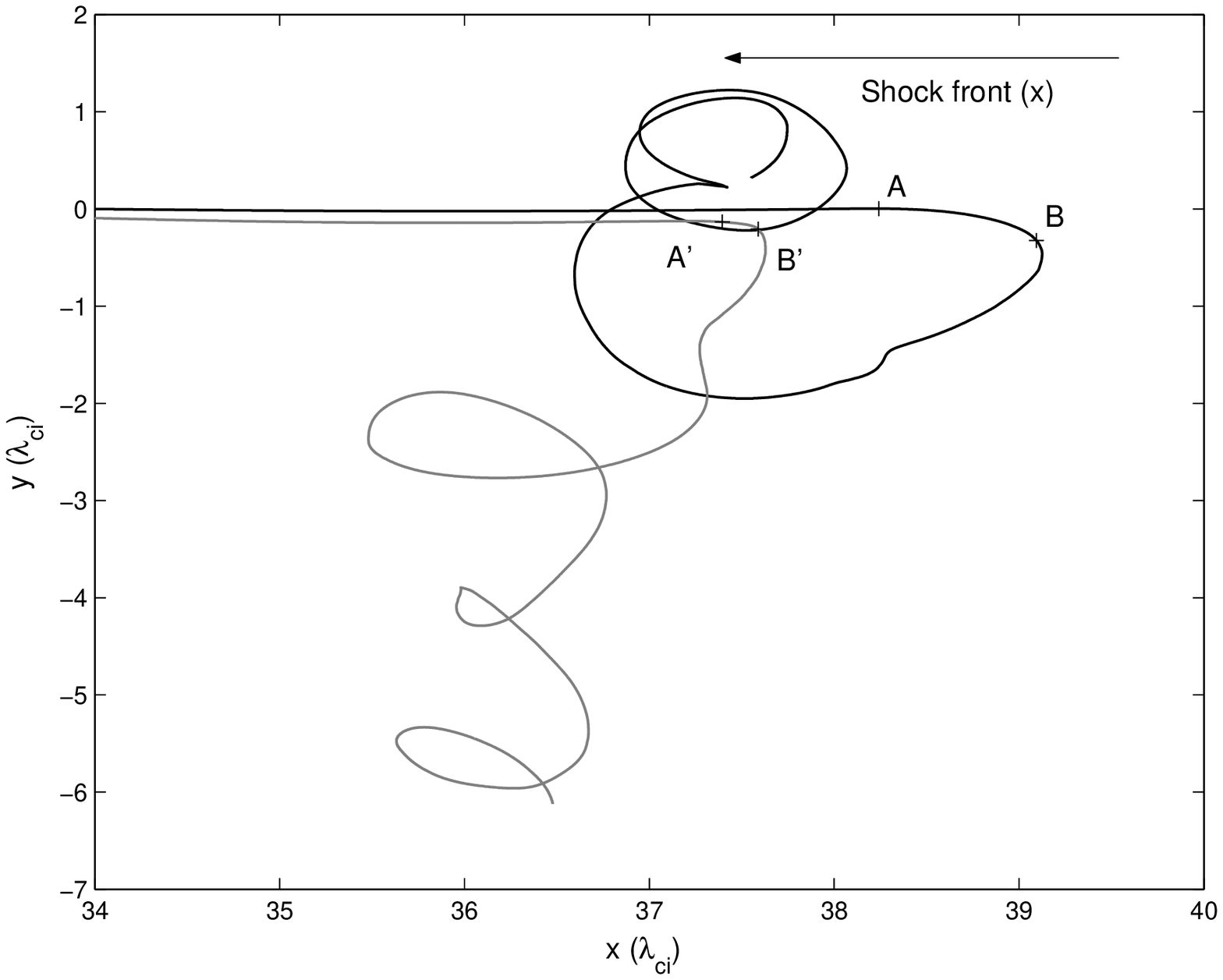}
\caption{Position $x$ vs. $y$ for two low energy ions. The black and
  gray trajectories from Fig.~\ref{fig:LowE-fdotv} are represented here,
  with the timing points $A,B,A',B'$ indicated.}
\end{figure}

\begin{figure}[floatfix]
\noindent\includegraphics[width=\columnwidth]{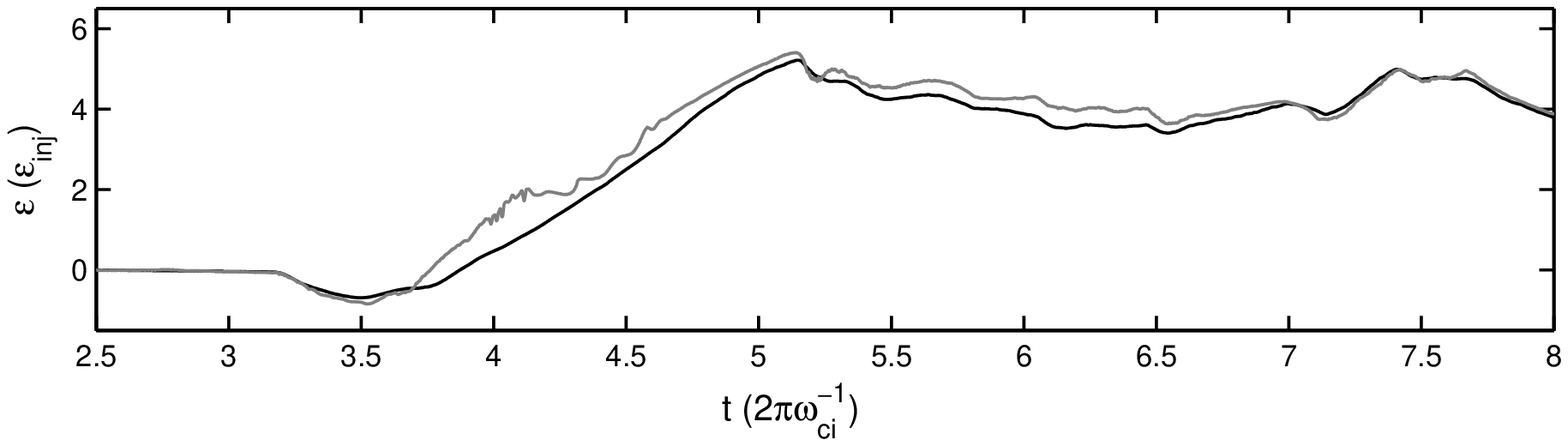}
\caption{Kinetic energy calculated from Eq.~(\ref{eqn:ke}) using the
  PIC simulation $\mathbf{E}$ (gray) and that on ion scales $\mathbf{E}_i$
  (black) from Eqs.~(\ref{eqn:ex}) and (\ref{eqn:ey}), for
  the black ion in Figs.~\ref{fig:HighE-fdotv} and \ref{fig:1p-xy}. The bulk
  ion velocity $v_{ix,y}$ in Eqs.~(\ref{eqn:ex}) and (\ref{eqn:ey}) is
  calculated from the mean velocity of the ions within $10$ grid cells
  $\simeq 0.02\lambda_{ci} \simeq \lambda_{ce}/2$ of the particle position.}
\end{figure}

\begin{figure}[floatfix]
\noindent\includegraphics[width =\columnwidth]{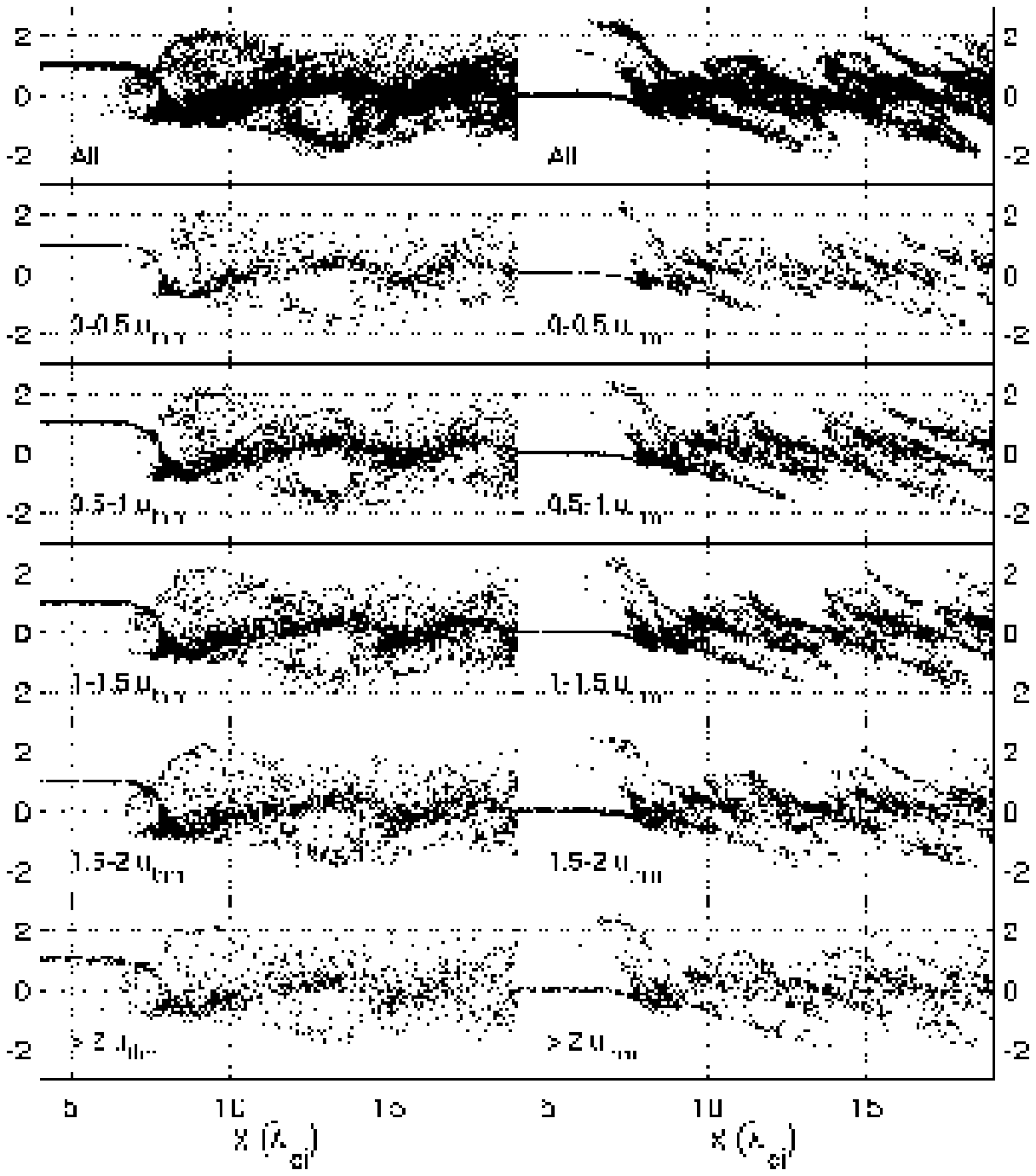}
\caption{Phase space plots of $v_x$ (left) and $v_y$ (right) vs. $x$ for
  groups of particles 
  from differing regions of initial velocity space, at the instant $t = 4
  \times 2 \pi \omega_{ci}^{-1}$.} 
\end{figure}

\end{document}